\documentclass[nofootinbib,a4paper,aps,preprint,twocolumn,preprintnumbers,amsmath,amssymb,10pt, superscriptaddress,prr]{revtex4-2}
\usepackage{graphicx}
\usepackage{color}
\usepackage[colorlinks=true,citecolor=blue,linkcolor=blue,urlcolor=blue]{hyperref}
\usepackage{braket}

\def\bbI{\mathbb{I}}
\def\bbO{\mathcal{O}}
\def\bbQ{\mathcal{Q}}
\def\TM{\mathcal{T}}
\def\tmE{\mathbb{E}^T}
\def\tmEp{\mathbb{E}^{T+1}}

\def\rank{\mathcal{R}}
\def\Lt{\Lambda^{s}}  
\def\Lb{\Lambda^{o}}
\def\Lbt{{\bar\Lambda}^{s}}  

\graphicspath{{./}{./pictures/}}

\begin{document}

 \title{On temporal entropy and the complexity of computing the expectation value of local operators after a quench.}
\author{Stefano Carignano}
\affiliation{Barcelona Supercomputing Center}
\email{stefano.carignano@bsc.es}

\author{Carlos Ramos Marim\'on}
\affiliation{Departament de F\'{\i}sica Qu\`antica i Astrof\'{\i}sica and Institut de Ci\`encies del Cosmos (ICCUB), Universitat de Barcelona,  Mart\'{\i} i Franqu\`es 1, 08028 Barcelona, Catalonia, Spain}
\author{Luca Tagliacozzo}
\affiliation{Institute of Fundamental Physics IFF-CSIC, Calle Serrano 113b, Madrid 28006, Spain}
\email{luca.tagliacozzo@iff.csic.es}

\begin{abstract}

We study the computational complexity of simulating the time-dependent expectation value of a local operator in a one-dimensional quantum system by using temporal matrix product states, and argue that it is intimately related to that of encoding temporal transition matrices and their partial traces. In particular, we show that we can upper-bound the rank of these reduced transition matrices by the one of the Heisenberg evolution of local operators, thus making connection between two apparently different quantities, the temporal entanglement  and the local operator entanglement.
As a result, whenever the local operator entanglement grows slower than linearly in time, we show that computing time-dependent expectation values of local operators using temporal matrix product states is likely advantageous with respect to computing the same quantities using standard matrix product states techniques.

\end{abstract}

\maketitle

\section{Introduction.}

The complexity of simulating quantum-many body systems increases exponentially with the number of its constituents. 
Over the last decades, the development of tensor networks (TN) techniques has however helped in gaining better insight on the equilibrium properties of many-body quantum systems. It is now understood that, at equilibrium, quantum complexity is mostly related to the amount of entanglement, and we have been able to design TN Ans\"atze that can encode the structure of typical equilibrium states. 

Out of equilibrium, the situation is different: even in the simplest protocol, such as e.g.\ quantum quenches,
 correlations spread over large distances, quickly producing robustly entangled states \cite{calabrese2005}. 
 Standard TN techniques such as those based on matrix product states (MPS) thus struggle to cope with the fast growth of entanglement and, as a result, their cost increases exponentially with the duration of the evolution \cite{laeuchli2008}: this is often referred to as “entanglement barrier” in the literature \cite{dubail2017}.

This fact is a consequence of trying to represent the full quantum state during the evolution. 
The situation might change if we focus on a local description of the state,  by trying to describe the evolution of the expectation value of local operators.
In principle, this is a much simpler task, 
and several approaches along these lines 
 have been proposed~\cite{banuls2009,muller-hermes2012,hastings2015,strathearn2018,white2017,rakovszky2020,frias-perez2022,lerose2021,lerose2023}. However, except for few studies on integrable systems~\cite{giudice2021,thoenniss2023}, there has been no systematic understanding of the real computational cost of such approaches and thus no concrete understanding on the complexity of simulating the evolution of local observables.

In this work we make a step in this direction. We consider the evolution under a local Hamiltonian of the expectation value of a local observable and follow the inspirational papers \cite{banuls2009,muller-hermes2012,hastings2015,frias-perez2022,lerose2023}. These algorithms rely still on a matrix product state, which is however now defined in time, making it a {\it temporal} MPS (tMPS) along a Keldysh contour \cite{hastings2015,tirrito2022}.
It has recently been noticed that such tMPS encodes the influence matrix of the system~\cite{feynman1963} that drives the evolution of a region of the full system, and thus provides the systematic way of translating the global evolution into a local one \cite{lerose2021,sonner2021,ye2021}.
 In specific cases, such tMPS can be described with small bond dimension,
 though this is not always the case  \cite{banuls2009,muller-hermes2012}.

Here we provide some theoretical backup to these numerical observations.
Our main result in this direction is to
bound
 the complexity of the tMPS with the one of encoding the time evolution of operators in the Heisenberg picture.
Such complexity is encoded in the so called ``operator entanglement'' (OE) \cite{prosen2007,pizorn2009},
which is expected to grow at most logarithmically in time for integrable systems, whereas it increases linearly for non-integrable systems~\cite{dubail2017,alba2019,bertini2020,bertini2020a}.
{The key to obtaining this result is a slightly modified version of the algorithms introduced in~\cite{banuls2009,hastings2015}, which, as shown in the following, allows to build a direct connection between the two quantities. }

Whenever the OE grows logarithmically in time, we can thus show that the tMPS bond dimension is bounded by a polynomial growth in time. On the other hand, when the OE grows linearly in time, our bound on the tMPS bond dimension is exponential.
While this does not rule out that there might be specific cases in which the tMPS might still have a small bond dimension and thus provide a computational advantage, see e.g.~\cite{piroli2020,bertini2020,bertini2020a}, we believe that these are but fortuitous exceptions to the general rule, and indeed in the scenarios we presented for this case as well as others in the literature \cite{frias-perez2022} we find that such bound is saturated.

\section{Setup}
\begin{figure}
  \includegraphics[width=\columnwidth]{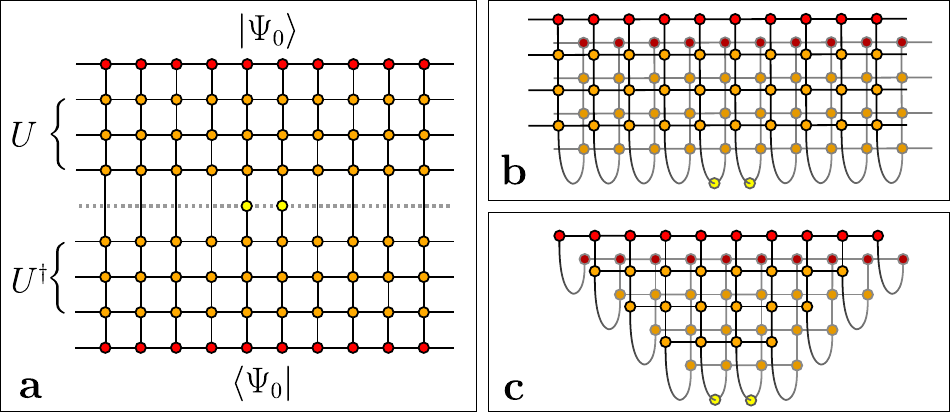}
 \caption{{\bf{(a)}} The time-dependent expectation value of a local operator acting on a one-dimensional system in the Keldysh representation 
 {\bf{(b)}} can be described by a double sheet 2D TN contraction. {\bf{(c)}} Upon using a Trotter approximation and the locality of the operator, it simplifies to a triangular TN.
 \label{fig:double_layer}}
\end{figure}

Given a lattice system, we consider the complexity of computing the evolution of the expectation value of a local operator after a quench.
We start from a product state $\ket{\psi_0}$ and a local Hamiltonian $H$. The system starts to evolve and it is described by the state $\ket{\psi(T)} = \exp{(-iH T)}\ket{\psi_0} \equiv U(T)\ket{\psi_0} $.
 In general, the entanglement entropy of the state increases linearly in time \cite{calabrese2005} and standard time-dependent MPS simulations become exponentially expensive \cite{vidal2004,white2004} (for recent reviews see also \cite{paeckel2019,banuls2023}). 
 Here, rather than evolving the state, we focus on the evolution of the expectation value of a local operator acting on two neighbouring sites \footnote{This can be easily generalized to an arbitrary number of sites as long as it is finite.},
\begin{equation}
 \braket{\mathcal{OQ}}(T) = \bra{\psi(T)}\mathcal{O}_i\mathcal{Q}_{i+1}\ket{\psi(T)}.
 \label{eq:loc_exp_val}
\end{equation}
Approximating  $U(T)$ by a sequence of short evolutions $U(\delta t)^{N_T}$ with $N_T= T/\delta t$ and using a Trotter expansion, this quantity is encoded in the contraction of  a two dimensional TN containing order $N_x\times N_T$ tensors \cite{mcculloch2007,pirvu2010}, see Fig. \ref{fig:double_layer}(a).
Furthermore, one can fold the network following  \cite{banuls2009,muller-hermes2012}, leading to a double-layer structure as shown in Fig.~\ref{fig:double_layer}~(b).
Given the Trotter approximation, the time evolution of a local operator has an exact causal cone obtained by cancelling all the unitary gates that are contracted with their respective Hermitian conjugates. As a result, the network can be simplified and acquires the triangular shape shown in Fig.~\ref{fig:double_layer}~(c) \cite{frias-perez2022,lerose2023}.

%
\subsection{Compressing the tMPS}
\begin{figure}
  \includegraphics[width=\columnwidth]{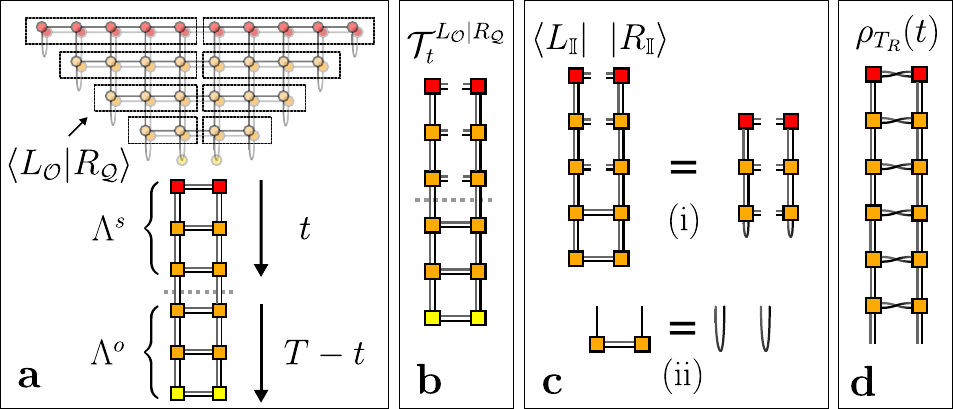}
\caption{ {\bf{(a)}}: The triangular TN can be contracted from the sides, identifying a left $\bra{L_{\mathcal{O}}}$ and a right $\ket{R_\bbQ}$ tMPS. 
The upper-most tensors of the temporal MPS are dictated by the initial state, while the lower-most ones by the choice of the operator.
{\bf{(b)}} The reduced transition matrix.
{\bf{(c)}} In the absence of operators, the folded tensors of the time evolution resolve to identities.
{\bf{(d)}} The partial transpose of the time-evolved left-right density matrix of the system, where forwards and backwards legs are swapped.
\label{fig:mps}}
\end{figure}

Being two-dimensional, the best contraction path for this TN is not a priori obvious.
a possible strategy is to identify two tMPS defining the contraction of the left and the right half of the system (see Fig.\ref{fig:mps}), and interpret the triangular network as the scalar product between the two, $
 \braket{\mathcal{OQ}}(t)  =\braket{L_{\mathcal{O}}|R_{\mathcal{Q}}}
$~\footnote{
Notice that this definition differs slightly from those of previous works, since the specific operators $\mathcal{OQ}$ are included in the definition of the tMPS.
By construction, if we chose as two-site operators the identity operator,  $\mathcal{OQ}=\mathbb{I}$ we obtain that
$
\braket{L_{\mathbb{I}}|R_{\mathbb{I}}} = \braket{\psi(t) | \psi(t) } = 1  \,.
$}.
The identification is purely formal, since the bond dimension of the individual tMPS tensors can grow exponentially with the number of time steps.
 The construction of the tMPS thus only makes sense if one can show that, at least for specific scenarios, its bond dimension increases mildly with the number of time steps.

%

The task is therefore to identify the relevant rank of the tMPS matrices and compress them on their support. We know that
the tMPS allows to compute correlation functions of the type $\langle \mathcal{OP}(t)\cdots\mathcal{QR}(t')\rangle$ for an arbitrary number of insertions of local operators at different times between $0$ and $T$.  Following the standard DMRG recipe \cite{white2004}, this requires having a faithful representation of all reduced transition matrices (RTM) $\mathcal{T}^{L_{\mathcal{O}}|R_{\mathcal{Q}}}_t$, defined as
 \begin{equation}
 \mathcal{T}^{L_{\mathcal{O}}|R_{\mathcal{Q}}}_t = \textrm{tr}_{T-t} \left[ \mathcal{T}^{L_{\mathcal{O}}|R_{\mathcal{Q}}}\right]  \,,\quad
 \mathcal{T}^{L_{\bbO}|R_{\bbQ}} =\frac{\ket{R_{\mathcal{Q}}}\bra{L_{\mathcal{O}}}}{\braket{L_{\mathcal{O}}|R_{\mathcal{Q}}}}\,,
  \label{eq:reduced_transition}
\end{equation}
which can be seen as generalizations of reduced density matrices. 

The graphical representation of these objects is found in~Fig.~\ref{fig:mps}(b).
Interestingly, such transition matrices have also been considered in the context of holography:
the properties of the equivalent in the field theory of the transition matrices described here have a geometrical interpretation in the bulk \cite{nakata2021,doi2023,doi2023,narayan2023,jiang2023}.
It is then the rank of such RTMs which dictates the rank of the tMPS matrices, rather than the temporal entropy that was previously considered~\cite{banuls2009,hastings2015,frias-perez2022,lerose2023}.

 Generalizing the DMRG prescription, the bond dimension of the  tMPS tensors
 should allow to obtain a low-rank approximation of the reduced transition matrix $\mathcal{T}^{L_{\mathcal{O}}|R_{\mathcal{Q}}}_t$, $\forall t \in 0\cdots T$, with the desired precision. 

The reduced transition matrices are complex-valued and,
for reflection invariant Hamiltonians,
whenever $\mathcal{Q}=\mathcal{O}$ they are symmetric.
As a result, their low rank approximation is better defined in terms of their singular values, so  we will project the tMPS bond dimension at time $t$ on the largest singular values of
$\mathcal{T}^{L_{\mathcal{O}}|R_{\mathcal{Q}}}_t$ \,.

Following \cite{hastings2015},
we now define $\Lt(t)$ and $\Lb(t)$ respectively as the contractions of the overlap $\braket{L_{\bbO}|R_{\bbQ}}$  until $t$ (the top part of the network contraction, including the initial state) and below $t$ (the bottom part, containing the operator), respectively, see Fig.~\ref{fig:mps}.
We also define
$\Lbt_L(t)$, $\Lbt_R(t)$ as the contraction up to $t$ of the network obtained by contracting $\braket{L_{\mathcal{O}}|L_{\mathcal{O}}}$ and $\braket{R_{\mathcal{Q}}|R_{\mathcal{Q}}}$ \footnote{For reflection-invariant Hamiltonians, $\ket{L_\bbO}$ differs from $\ket{R_\bbO}$ only by a complex conjugation}.

With these definitions, we have that
\begin{equation}
 \mathcal{T}^{L_{\mathcal{O}}|R_{\mathcal{Q}}}_t \simeq \sqrt{\Lbt_L(t)}\Lb(t)\sqrt{\Lbt_R(t)} \,,
 \label{eq:relation_between_entr}
 \end{equation}
where we use the similarity symbol to indicate that the two operators share the same singular values.
For reflection invariant systems, such as those we analyze here, one has $\Lbt_R(t)=\Lbt_L(t)$, thus our central result reads
\begin{equation}
\rank\left(\mathcal{T}^{L_{\mathcal{O}}|R_{\mathcal{Q}}}_t\right) \le \min \left\{ \rank\Big(\Lbt_L(t)\Big),\rank\Big(\Lb(t)\Big)\right\} \,,
 \label{eq:central_result}
 \end{equation}
where $\rank$ denotes the rank of a given matrix.
This equation is at the basis of our analysis. 

Since for each choice of  $\bbO$ and $\bbQ$ we obtain a different tMPS representation or $\bra{L_{\mathcal{O}}(t)}$ and $\ket{R_{\mathcal{Q}}(t)}$,
we define the cost of the algorithm as 
the cost of simulating the operators that require the highest bond dimension.

\subsection{The rank of the reduced transition matrices}

 Whenever $\mathcal{O}= \mathcal{Q} = \mathbb{I}$, the transition matrices $\mathcal{T}^{L_{\mathbb{I}}|R_{\mathbb{I}}}_t$ for all $t$ are  projectors. As a result the states
$\ket{L_{\mathbb{I}}}$ and $\ket{R_{\mathbb{I}}}$ consist of trivial singlets at the virtual level, as sketched in Fig.~\ref{fig:mps} (c), implying that the tMPS matrices have always bond dimension 1, for all $t$.
In order to obtain non-trivial tMPS we thus include the operators in the construction of the states.

From the previous definitions it follows that
\begin{equation}
 \Lb(t) = U(t)\left( \mathcal{O}\otimes\mathcal{Q}\right) U(t)^{\dagger} \,,
 \label{eq:heis}
\end{equation}
meaning that $\Lb(t)$ exactly encodes 
 the Heisenberg evolution of the initially localized operator $\mathcal{O}_i \mathcal{Q}_{i+1}$, while 
\begin{equation}
 \Lt(t) = \rho(t) = U(t)  \ket{\psi_0}\bra{\psi_0} U(t)^{\dagger}  \,,
 \label{eq:evolved_state}
\end{equation}
where $\rho(t)$ is the time-evolved density matrix of the initial state.
Finally, for Hamiltonians that are invariant under  reflections with respect to the center of the chain,
\begin{equation}
\Lbt_L(t) = \Lbt_R(t) = \rho(t)_{T_R} \,, \label{eq:partial_transpose}
\end{equation}
where  $T_R$ stands for the partial transpose on the semi-infinite right part of the system.
 Since we are evolving a pure state, 
there are known relations between $\rho_{T_R}(t)$ and $\rho(t)$ \cite{coser2014},
and in particular we know that $\rank \left( \Lbt_{R}(t) \right) =\rank \left( \Lt_R(t) \right) $ \cite{calabrese2012, vidal2001}. 

Having identified the elements composing the RTMs in Eq.~\eqref{eq:relation_between_entr}, we can now use their physical properties together with Eq.~\eqref{eq:central_result} to identify some useful bound on the rank of the tMPS. 
A first scenario one can consider is that of a local quench. In this case, it is well known that the entropy of the evolved state only increases logarithmically with time, and as a result the rank of $\Lt(t)$ increases at most polynomially \cite{calabrese2007,verstraete2010}, thus using Eq.~\eqref{eq:central_result} we have
\begin{equation}
 \rank \left(\mathcal{T}^{L_{\mathcal{O}}|R_{\mathcal{Q}}}_t\right) \le T^{\alpha} \quad \forall t \in \set{0,T}.\label{eq:res_loc_q}
 \end{equation}
 Such a case is perhaps not particularly interesting, since a similar polynomial cost is obtained also by using standard MPS algorithms.

 We thus turn to the more interesting scenario of a global quench. Here we know that $\rank\left(\Lt(t) \right)$ increases exponentially with $T$, as already mentioned.
   If now $\rank \left(\Lb(t)\right)$ only increases polynomially with $T$, we have
\begin{equation}
 \rank \Big( \Lb(t) \Big) \le t^{\alpha} \, \Rightarrow \, \rank \left( \mathcal{T}^{L_\bbO |R_\bbQ}_t \right) \le T^{\alpha} \quad \forall t \,,
\label{eq:main_bound}
\end{equation}
and the temporal MPS strategy provides a polynomial algorithm to compute the out-of-equilibrium dynamics of local observables, as already anticipated in \cite{frias-perez2022,lerose2023}.
 In the literature, it is conjectured (and checked in several scenarios) that in the case of integrable systems the entanglement of local operators only grows logarithmically \cite{prosen2007,pizorn2009,dubail2017,bertini2020,bertini2020a}, implying that the rank of $\Lb(t)$ only increases polynomially with $T$ as required in Eq.~\eqref{eq:main_bound}. As a result, if the conjecture is correct, the tMPS provides an efficient method to characterise the time evolution of local operators for integrable systems.

Since generically  the rank of $\Lb(t)$ is expected to grow exponentially with $t$, we expect that
\begin{equation}
\rank \left(\mathcal{T}^{L_{\mathcal{O}}|R_{\mathcal{Q}}}_{t} \right) \le  \alpha \exp(\beta) \,,
\end{equation}
with $\alpha$ and $\beta >0$ model-dependent constants.
In this case, there is no guarantee that the tMPS provides an efficient compression for the problem at hand.

\subsection{Relation to other definitions of temporal entanglement}

The basic ingredient in our formalism are the transition matrices defined in Eq.~\ref{eq:reduced_transition}, and the generalized entanglement related to their rank. 
 It is worth noting that these quantities differ from the definition of temporal entanglement recently proposed in the literature~\cite{banuls2009,hastings2015,frias-perez2022,lerose2023,giudice2021}, which is built through density matrices obtained individually from either the left or right vectors (ie. $
\rho_R \sim \ket{R}\bra{R}, \rho_L \sim \ket{L}\bra{L}$).

This can be most easily seen in the case of a reflection-symmetric problem: here one has $\ket{R} = \ket{\bar{L}}$, so that the two definitions are connected by a nontrivial complex conjugation on the left state, which can be seen as a partial transpose operation acting on the legs of the folded transition matrix (see Fig.~\ref{fig:mps} (d)).
This global unitary operation (the tensor product of swap gates on every physical leg, ie. at each time site) acting on one side of the transition matrix can change the entanglement properties of the objects involved, so that a clear connection between the two quantities is not evident and would definitely require a deeper investigation in the future.

Another aspect which has been recently pointed out is that
 the traditional definition of temporal entanglement can depend on the gauge chosen to define the transverse transfer matrix, and can be made arbitrarily small by using such gauge freedom. As a result, being able to accurately represent $\ket{R}$  and $\bra{L}$ does not guarantee obtaining a good precision on an expectation value  $\braket{O(T)} = \bra{L}O\ket{R}$, since the leading Schmidt vectors can be orthogonal to such overlap  \cite{tang2023}.
Thus, as we argued, for our intent of determining the complexity of simulating expectation value of local operators, the use of the generalized entanglement proposed here is the appropriate object to consider, close in spirit to the concept of biorthonormal truncation basis, which is already well-established for non-hermitian DMRG problems \cite{wang1997,huang2011,tang2023}.

\section{Numerical results.}
We now report the numerical evidence to support our claims. We consider a transverse field Ising Hamiltonian
\begin{equation}
{{H}}(g, h) = -  \sum_i  \Big[ \sigma_x^i \sigma_x^{i+1} + g \sigma_z^i + h \sigma_x^i \Big] \,,
\end{equation}
where $\sigma_{x,z}$ are Pauli matrices. 
We perform a second-order Trotter expansion of the time evolution operator and cast it into a matrix product operator (MPO) following \cite{pirvu2010}, with a timestep $\delta t=0.05$.  
For our examples we consider two cases: an integrable one for $g=0.5, h=0$, and a non-integrable one $g=-1.05, h=0.5$. 
We consider different initial product states, namely $\ket{0} = (1,0)^{\otimes N_x }$, $\ket{+} = (1,1)/\sqrt{2}^{\otimes N_x }$ and $\ket{r} =  (1,i)/\sqrt{2}^{\otimes N_x }$ . The time evolution is performed by the variant of the standard folding algorithms we have described, exploiting the causal-cone of the network \cite{banuls2009,hastings2015,frias-perez2022,lerose2023} and relying on a low-rank approximation of the RTMs defined in Eq.~\eqref{eq:reduced_transition}.

As in previous approaches \cite{frias-perez2022,lerose2023}  we iteratively construct the left and right vectors     $\bra{L_{\mathcal{O}}(T)}$ and $\ket{R_{\mathcal{Q}}(T)}$ starting from $\bra{L_{\mathcal{O}}(T-1)}$ and $\ket{R_{\mathcal{Q}}(T-1)}$  by absorbing a new column of MPOs into them.
These MPOs are obtained by contracting a column of the original TN and they thus represent transfer matrices $\tmE$ evolving states of a time-slice in space, a {\it{rotated}} version of the standard evolution in time.
The bond dimension of the corresponding MPS increases, and we then keep it under control by projecting onto the largest $D$ singular values of $\mathcal{T}^{L_{\mathcal{O}}|R_{\mathcal{Q}}}_t$ for every bi-partition of the system into $t$ and $T-t$. Further details about the algorithm and accurate comparison with other prescriptions in the literature are presented in the accompanying Supplementary Material.

 In the case of the Ising model, we found the operator with the largest bond dimension defined at most on two sites to be a single-site $\sigma_x$. In the following we will focus on optimizing with respect to it \footnote{For operators with known small bond dimension such as $\sigma_z$ or $\sigma_x \sigma_x$, our algorithm saturates at the expected fixed value of $D$  \cite{hartmann2009}.}.

\begin{figure}
\includegraphics[width=.8\columnwidth]{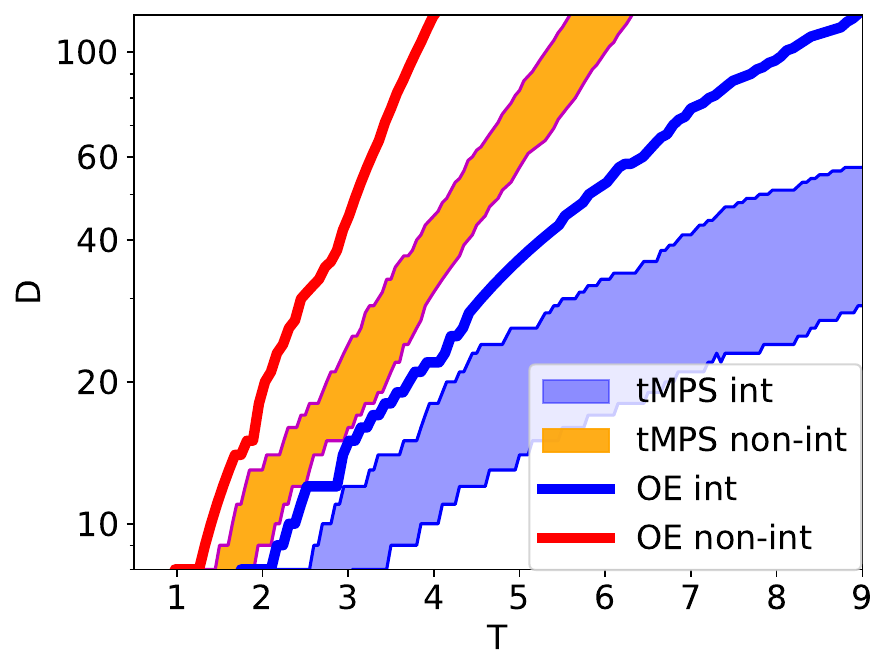}
\caption{ Left: Bond dimensions $D$ obtained by imposing a truncation error of $10^{-4}$ in the singular value decomposition of the RTMs, as function of time (see main text). The shaded areas are delimited by the minimum and maximum $D$ we found by varying the initial state.
 The thick solid lines represent the bond dimension curves for the operator entanglement (OE): they lie consistently above the corresponding ones for the tMPS.
 Blue colors denotes the integrable case (int), orange-red the non-integrable one (non-int).}
\label{fig:epscut}
\end{figure}

We start by considering the bond dimension necessary to keep the truncation error above a threshold $\epsilon = 10^{-4}$ in the SVD spectrum of the reduced transition matrices, and compare them with those obtained imposing the same truncation on 
the Heisenberg evolution for the operator $O(t)$, which entails the evolution of the vectorized local operator under the Hamiltonian $H\otimes \mathbb{I} -\mathbb{I} \otimes H$, resulting in the  MPS, $\ket{\psi_O(t)}$.
 The results are shown in Fig.~\ref{fig:epscut}, where we can observe that, as expected: 1) the behavior is different in the integrable and non-integrable cases, 2) the bond dimension necessary to correctly describe the RTM are always below the ones necessary to describe the Heisenberg evolution of the operator.

We note however that in order to estimate the computational cost of simulating the expectation value of local operators with a finite precision using a tMPS we should determine the bond dimension that is required to keep  the distance in norm $|| \bra{L_\bbO} - \bra{L_\bbO^D}||^2 $ constant, where $\bra{L_\bbO^D}$ is the
truncation of the tMPS $\bra{L_\bbO}$ to a given bond dimension $D$. Using the reasoning in Ref.~\cite{verstraete2010}, the term to consider would then be the overlap  $\braket{L_\bbO | L_\bbO^D}$, which is upper bounded by the sum of the norm of the residuals discarded in the truncation process.

So far we have just provided a bound on the overlap $\braket{L_\bbO | R_\bbO}$ which, even for reflection-invariant systems and considering the symmetric case $\bbO = \bbQ$, differs from $\braket{L_\bbO | L_\bbO^D}$  by the absence of a complex conjugation. However, the two quantities are related, since  $\braket{L_\bbO | L_\bbO^D}$  can be obtained from $\braket{L_\bbO | R_\bbO}$ by evolving   $\ket{R_\bbO}$ with one layer of local unitaries, representing the partial transpose (swapping the bra with the ket legs). 
Such an operation does not modify the rank of each state, but it reduces the overlap between the two, so that a shift in the resulting singular values can be expected.

As a result, the computational cost required for simulating the expectation value of a local operator with a finite precision is constantly offset from that of computing $\braket{L_\bbO | R_\bbO}$, the latter being upper-bounded by the cost of the Heisenberg evolution of the operator. The scaling in time of the cost to maintain a constant error in local observables
is the same as the one of  $\braket{L_\bbO | R_\bbO}$ (see also the discussion in the Supplementary Material).

\begin{figure}
\includegraphics[width=.8\columnwidth]{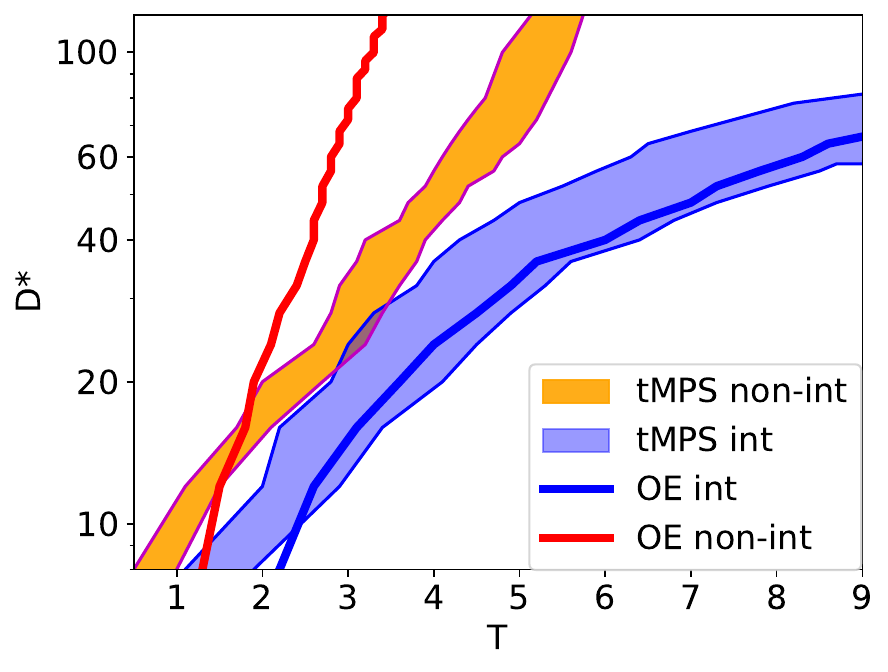}
\caption{ Bond dimensions $D^*$ obtained by imposing a $10^{-4}$ fidelity for the tMPS $\bra{L_\bbO}$, again the shaded area is delimited by the minimum and maximum bond dimension of the tMPS for a given initial state, and the thick solid lines denote the operator entanglement (OE) (blue: integrable case, red: non-integrable)}
\label{fig:dstars}
\end{figure}

We verify this explicitly by following a procedure similar to the one proposed in \cite{prosen2007} to estimate the truncation error in our tMPS compared to an exact result. The resulting bond dimensions for a fixed truncation error are reported in Fig.~\ref{fig:dstars}. For the integrable case, we find a polynomial increase, consistent with the behavior of the OE. Different initial states require different bond dimensions for a faithful description, though the growth follows the same power law. 
For the non-integrable case, on the other hand, the OE requires an exponentially growing bond dimension for a faithful description, and the tMPS bond dimension follows the same behavior, consistent with our result.

\section{Conclusions.}
We have provided a new connection between two separate concepts, temporal entanglement and the operator entanglement. In particular, we have shown that the rank of the RTM necessary to compute the temporal entanglement is upper-bounded by the rank of the OE of a bi-partition. As a result, the  scaling in time of the computational cost of simulating the evolution of a local observable with constant error is upper-bounded by the scaling of the OE, even though the exact value of the cost can be offset by a constant value whose origin has been discussed in detail.
Our presented algorithm yields the most efficient performance observed thus far, with a marginal advantage over previous approaches.

As a result, we can claim that whenever the OE only grows logarithmically, using a tMPS for simulating the evolution of a local observable is the best choice in terms of computational cost and scales only polynomially with time, whereas in the generic case the entanglement barrier cannot be circumvented by using tMPS.
On the other hand, a linear growth of the OE does not necessarily imply that the tMPS cannot provide an efficient compression, see eg.~\cite{piroli2020,bertini2020,bertini2020a}. The transverse contraction methods discussed here would then be useful even in those cases.

Our results are a step towards understanding the cost of simulating the out-of-equilibrium dynamics of quantum many-body systems.
It would be interesting to check the bounds we have obtained on a larger class of integrable and non-integrable models, and more importantly to explore if the transverse contraction can help in simulating the dynamics of higher dimensional systems. 

While working on this paper, we found out about Ref.~\cite{foligno2023} discussing similar issues and suggesting a  connection between the OE and the temporal entanglement, a connection we hope to have elucidated with our work.

\begin{acknowledgments}
We acknowledge the precious discussions on this and related subjects with Mari-Carmen Bañuls, Miguel Fr\'ias-P\'erez, Dimitry Abanin, Toma{\v z} Prosen, Ignacio Cirac, Pavel Kos, Jan Schneider,  Georgios Styliaris and Wei Tang. LT acknowledges support from the Proyecto Sin\'ergico CAM 2020 Y2020/TCS-6545 (NanoQuCo-CM), the CSIC Research Platform on Quantum Technologies PTI-001 and from Spanish projects PID2021-127968NB-I00 and TED2021-130552B-C22 funded by MCIN/AEI/10.13039/501100011033/FEDER, UE and MCIN/AEI/10.13039/501100011033, respectively.
CR acknowledges support from a “la Caixa” Foundation fellowship (ID 100010434, code LCF/BQ/DI21/11860031).
This work has been financially supported by the Ministry of Economic Affairs and Digital Transformation of the Spanish Government through the QUANTUM ENIA project call – Quantum Spain project, and by the European Union through the Recovery, Transformation and Resilience Plan – NextGenerationEU within the framework of the Digital Spain 2026 Agenda.

\end{acknowledgments}

\section{Appendix}

\appendix

\subsection{The algorithm}

\noindent
In this Appendix we describe the algorithms we employ to obtain the left and right tMPS 
 encoding the contraction of the TN associated with the expectation value $\braket{\bbO(T)}$, where $\bbO$ is a local (or a set of local) operator acting on one or few neighbouring sites.
 In practice, for the iterative methods we employ it is convenient to incorporate the operator(s) on one side of the network, as will become clear in the following. If we choose eg. to incorporate $\bbO$ on the right side of the network, our algorithms allow to build $\bra{L_\bbI}$ and $\ket{R_\bbO}$ for a given time $T$.

The basic building blocks of the TN are the MPO tensors representing the trotterized time evolution (here we employ the parametrization from \cite{pirvu2010}). For a given $T$ which determines the extension of the network in the temporal direction ($N_T = T/\delta t$, $\delta t$ being the Trotter step), we contract them in columns to generate the  transfer matrices encoding the rotated (space-like) evolution $\tmE$ for the folded system of which  $\bra{L}$ and $\ket{R}$ are the leading left and right eigenvectors in the thermodynamic limit. Depending on what we want to construct, we can also include the local operator at the bottom of the transfer matrix, in that case we label it as $\tmE_\bbO$.

We have implemented both a power-method  for extracting the leading eigenvectors of $\tmE$ for a fixed time $T$, as well as an algorithm based on the strict causal cone of the network when dealing with local operators.  Both our algorithms are small variations of those presented in \cite{banuls2009,frias-perez2022,lerose2023}), the main difference being the cost-function we use to perform the truncation required in the various iterations: we project the tMPS matrices on the support of the largest singular values of the respective RTM, as will be explained in the following.

\begin{figure}
 	\includegraphics[width=\columnwidth]{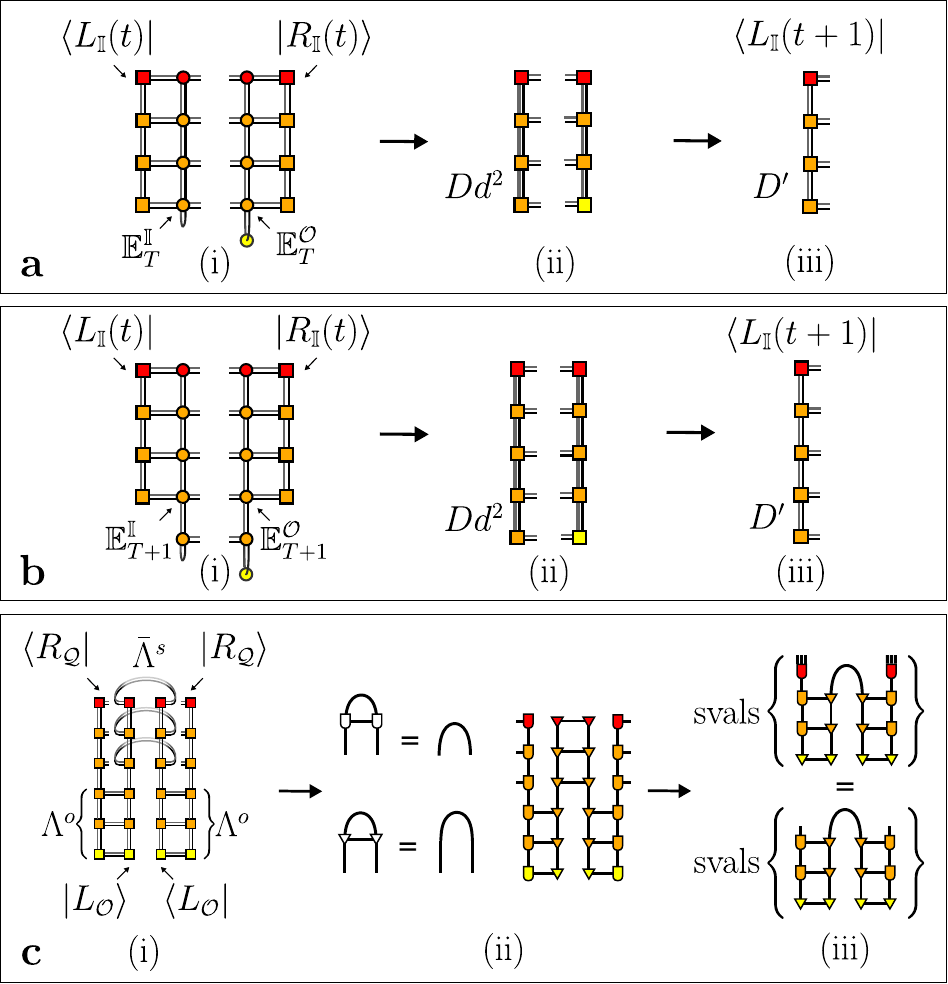}
 	\caption{ Illustrations of the methods used for building the tMPS. {\bf{(a)}} Power method {\bf{(b)}} Light cone method {\bf{(c)}} The matrix involved in our low-rank approximation of $\TM$ and the role of the gauge transformation in our calculation. }
 	\label{fig:methods}
 \end{figure}

The {\it power method}  (see Fig.~\ref{fig:methods}(a)) works by repeatedly applying the transfer matrix to an initial guess tMPS of length $N_T$ until convergence is reached. More specifically, we start from the left $\bra{L_\bbI}$ and right $\ket{R_\bbI}$ vectors, and apply a column $\tmE$ to the left and a  $\tmE_\bbO$ to the right. At each step, 
the bond dimension of the tMPS increases by a factor $d^2$, where $d$ is the physical dimension of the system constituents,
so in order to proceed we truncate following our prescription and take the updated $\bra{L_\bbI}$ as input for the next step. With our parametrization of the MPO tensors, $\tmE$ is symmetric in left-right legs, so we can use  $\bra{L_\bbI}$ also as the new $\ket{R_\bbI}$. Alternatively, the optimization for  $\ket{R_\bbI}$ simply requires an analogous step involving $\bra{L_\bbO}$.

In order to determine whether the power method has converged, we calculate several entropies associated with the  $\bra{L_\bbI}$ and $\ket{R_\bbO}$ vectors: most notably, the Von Neumann entropy $S_t^{VN}({L_\bbI})$ and the generalized R\'enyi 2, which we define as $ S^{r2}_t({R_\bbO}) = -\log \sum_n \lambda_n^2$, where $\lambda_n^2$ are the eigenvalues of the RTM $\TM^{L_\bbO|R_\bbO}_t$\footnote{This particular choice ensures that the RTM is symmetric, which guarantees that its diagonalization is not an ill-posed problem. }. Convergence is reached at a given step $i$ of the power method if, for the aforementioned entropies, one has $\sum_{t=0}^{N_T}  (S^{i-1}_t - S^i_t)^2 < \epsilon$, which in our case we take to be $\epsilon=10^{-6}$.

The {\it light-cone method} works in a slightly different way, as it allows to build the tMPS for a time $T$ starting from the one for $T-1$ with a single optimization (see Fig.~\ref{fig:methods}(b)): Starting from $\bra{L_\bbI}$ and $\ket{R_\bbI}$ of a given length $N_T$, we apply a transfer matrix $\tmEp$ of length $N_T+1$ to the left and a $\tmE_\bbO$ to the right, extending the network both in the time and space direction.

As for the power method, after each iteration the bond dimension of the tMPS grows by a factor $d^2$, so that truncating is required to avoid an exponential computational cost. The truncated $\bra{L_\bbI}$ and $\ket{R_\bbI}$ are then used as starting point for the next iteration.
The cone is thus built from the center moving outwards, assuming that the truncation of its edges does not spoil the causal structure induced by the local operator. The algorithm seems in any case to be quite stable, as we checked by piling up periodically a few layers of $\tmE$ before truncating and verifying that the final result is the same as the one obtained when we truncate after each iteration.

In spite of their differences, both methods end up giving comparable results, as they share the same truncation procedure, which goes as follows: We focus on optimizing the overlap between $\bra{L_\bbI}$ and $\ket{R_\bbO}$, which includes the operator $\bbO$ whose expectation value we are interested in.  Since our aim is to find a low-rank approximation for the RTM $\TM^{L_\bbI | R_\bbO}_t$ at any bi-partition $t$ and $T-t$, we focus on computing the eigenvalues of the matrix $(\TM^{L_\bbI | R_\bbO}_t)^\dagger \TM^{L_\bbI | R_\bbO}_t$ depicted in Fig.~\ref{fig:methods}(c), which can be associated with the squares of the relevant singular values. As usual in TN calculations, we can transform a global optimization problem into a local one by making use of gauge transformation \cite{orus2008,evenbly2022}. In particular, we start by bringing both $L$ and $R$ vectors individually to a standard orthogonal gauge, starting from the side of the initial state (Fig.~\ref{fig:methods}(c)). This gauge transformation allows to incorporate the  $\Lbt_{L,R}(t)$ factors towards the operator side, so that we don't need to compute them explicitly as their contraction up to $t < T$ now reduces to an identity. We now only need to contract the bottom environments and project them over their largest $D'$ singular values at each  $t$, thus optimizing the overlap between the two tMPS.
The advantage of keeping the local operators on either the left or the right side of the network is that, while we optimize with respect to $\braket\bbO$, we always end up with a new $\bra{L_\bbI}$ ($\ket{R_\bbI}$) which does not include the operator itself.  This tMPS, which is related rather to the time evolution of the state, thus encodes the influence functional of the system, and can be used as starting point for the next iteration of our algorithms.

\subsection{Estimate of the bond dimension for a given fidelity }
In order to estimate the bond dimension required to obtain our tMPS with a given fidelity, we follow a procedure similar to the one employed in \cite{prosen2007} to estimate the truncation error in our tMPS compared to an exact result, which in our case corresponds to the evolution obtained with the maximum bond dimension  we can afford, $D_{\max}( \simeq 1000$).
%
The largest $T$ we consider for this estimation are thus restricted to times for which $D_{\max}$  does not induce any sizeable truncation error with respect to the exact (Trotterized) dynamics. If we work with normalized MPS,  the error induced by truncating the operator MPS  to bond dimension $D$ is encoded in the fidelity $F = |\braket{\psi_O^{D_{\max}}(t)| \psi_O^D(t)}|$, where $\ket{\psi_O^D(t)}$ represents the time-evolved state up to time $t$ truncated at every time step to  bond dimension $D$. By fixing a given accuracy $\epsilon$,  we thus identify the maximum time for which  $F 
\ge 1-\epsilon$. As a consequence, by repeating the above procedure for different $D$ we can identify a curve $D^*(t)$ along which the truncation error is kept roughly constant at a value $\epsilon$.

\begin{figure}
	\includegraphics[width=\columnwidth]{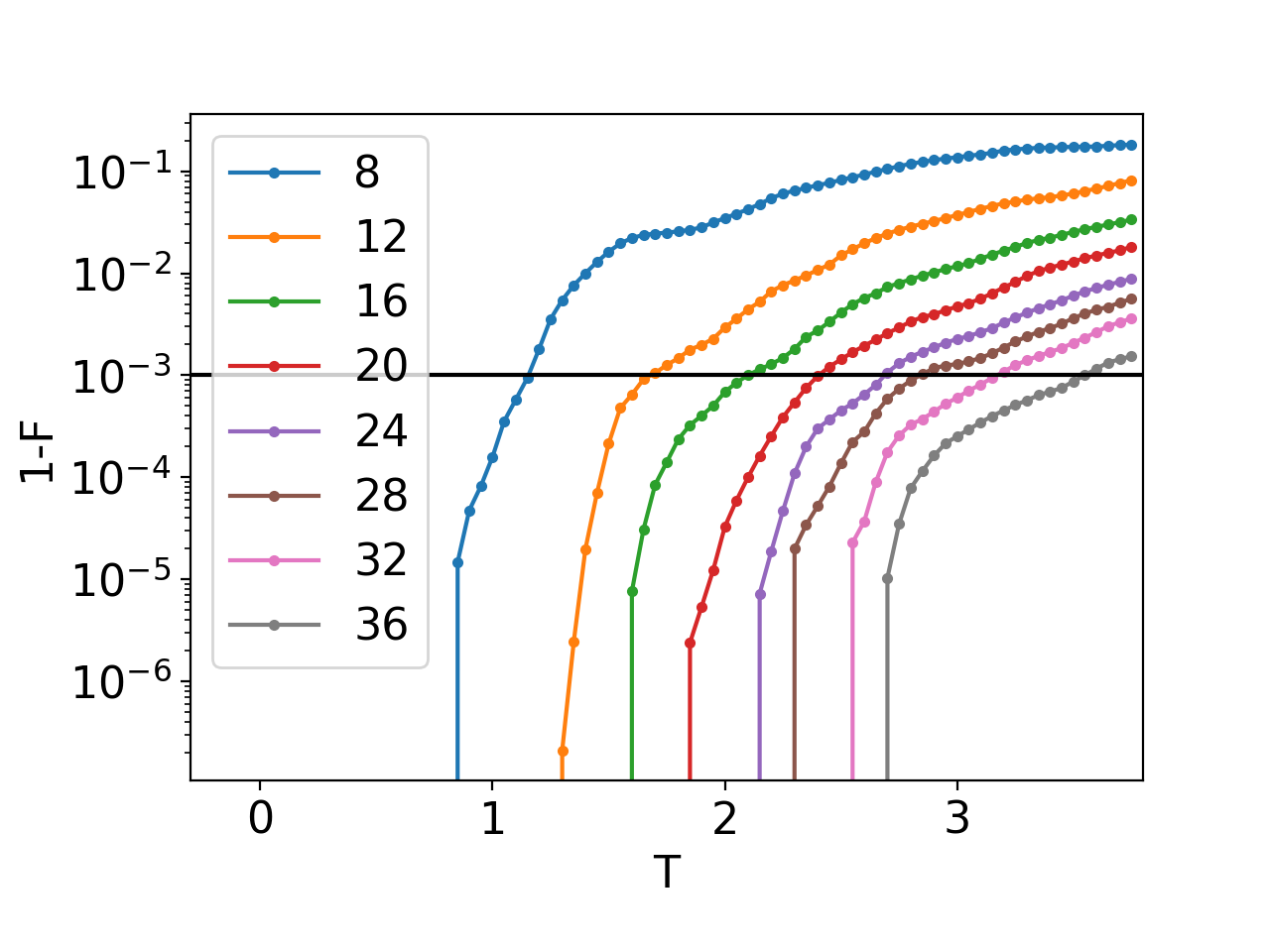}
	\caption{ Fidelity for overlap $\braket{L^{D_{\max}}_\bbO|L_\bbO^D}$ for different bond dimensions, as function of time, for the integrable case with starting state $\ket{+}$. After fixing a threshold $\epsilon$ we can extract the required bond dimension to give a faithful approximation of  $\ket{L_\bbO^D} $.
	\label{fig:fidelities}}
\end{figure}

We then extract a similar curve $D^*(t)$ for the tMPS encoding the semi-infinite left (right) TN $\bra{L_\bbO}(\ket{R_\bbO})$. The truncation here is based on projecting onto the largest singular values of each RTM, as described above. The largest  $D_{\max}$ we can afford is again used as the exact reference. The overlap of the two tMPS measures the effects of the truncation error $F=|\braket{L^{D_{\max}}_\bbO|L^D_\bbO}|$. Once more by fixing such an error to a given threshold $\epsilon$ we can identify the corresponding $D^*(t)$ for the tMPS. The typical behavior of the resulting curves for $F$ can be seen in Fig.~\ref{fig:fidelities}.

\subsection{Comparison with other approaches}

As previously mentioned, our optimization prescription is a relatively small modification with respect to already existing algorithms, which allows nevertheless to make a clearer connection with the underlying structures and estimate the computational complexity associated with the calculation of time-dependent expectation values. 
It might nevertheless be interesting to check how this method compares with other prescriptions employed in the literature.

 \begin{figure}
 	\includegraphics[width=\columnwidth]{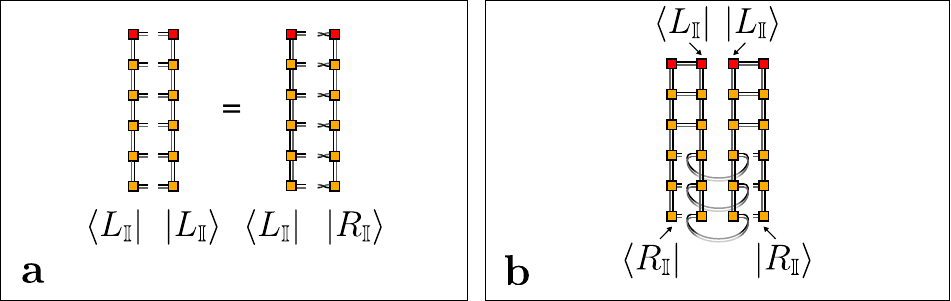}
 	\caption{ {\bf{(a)}} A possible method for truncating the tMPS relies on the optimization of the overlap $\braket{L|L}$, where in the symmetric case $\ket{L}$ can be related by a partial transpose to the vector $\ket{R}$. {\bf{(b)}} Alternatively, it is also possible to consider the RTM starting from the side of the initial state, a procedure which can give a nontrivial SVD spectrum even in the absence of a local operator.  }
 	\label{fig:other_methods}
 \end{figure}

The most straightforward optimization procedure when dealing with (t)MPS
after applying a column of MPO is to truncate using the standard canonical forms for the left and right vectors separately. In our framework, 
with left-right symmetry of the Hamiltonian and translational invariance, 
this would amount to an optimization of $\bra{L}$ with its conjugate $\ket{L}$ instead of $\ket{R}$. The operation of complex conjugation here can be seen as a partial transpose involving the forwards and backwards leg of the tMPS (basically an insertion of a series of swap operators), see Fig.~\ref{fig:other_methods} (a), since (not considering additional complications in case the initial state is complex) it basically amounts to an exchange $U(t) \leftrightarrow U^\dagger(t)$ of the time evolution operators. Due to the non-trivial structure induced by this, the projector structure of the TN is lost even if no operators are present, so that one obtains a non-trivial bond dimension even when $\bbO = \bbI$. 

Instead of optimizing with respect to a single operator inserted at the edge of the MPO column, the conjugation would imply acting on the whole column, ie. to perform these transpositions at each timestep.

Another possible strategy, which has been suggested in \cite{hastings2015,frias-perez2022} involves bringing the tMPS to canonical form starting from the bottom side and then doing a sweep from the initial state optimizing $\braket{L_\bbI | R_\bbI } $, Fig.~\ref{fig:other_methods} (b).
 In this case, while the spectrum of the RTM is trivial since no operator is present, the singular value decomposition is not: this can be seen again as due to an insertion of swap operators which generate a non-trivial spectrum, which however cannot be directly related to the causal structure generated by a local operator. In this case, one is unable to give a priori a statement on the maximum rank required by the algorithm.

\begin{figure}
	\includegraphics[width=\columnwidth]{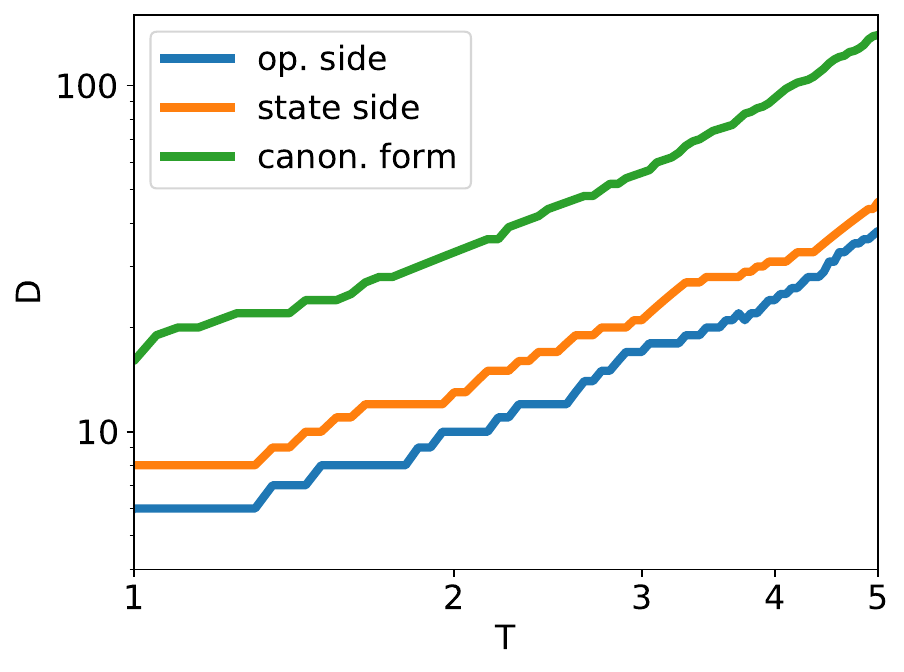}
	\caption{ Bond dimensions of the tMPS as function of time obtained with different methods (as described in the text): our method (blue line), the optimization starting from the side of the initial state (orange) and the canonical form (green), imposing the same truncation error ($\epsilon_{trunc} = 10^{-6}$). Our method leads to the smallest bond dimension. 
	\label{fig:chis_different_methods}
	}
\end{figure}

In spite of this difference, remarkably this latter method returns comparable bond dimensions to the ones obtained with our method when $\bbO = \sigma_x $, see Fig.~\ref{fig:chis_different_methods}. On the other hand, the optimization using the standard canonical forms requires a larger bond dimension, although the behavior of the various algorithms is comparable: this confirms our expectation that the scaling of our truncation based on the overlap $\braket{L|R}$ is the same  as the one based on $\braket{L|L}$.

%

\end{document}